# Evolution of proto-galaxy-clusters to their present form: theory and observations

Carl H. Gibson [1,2]

[1] University of California San Diego, La Jolla, CA 92093-0411, USA
[2] cgibson@ucsd.edu, http://sdcc3.ucsd.edu/~ir118

and

Rudolph E. Schild [3,4]

[3] Center for Astrophysics, 60 Garden Street, Cambridge, MA 02138, USA
[4] rschild@cfa.harvard.edu

## ABSTRACT

From hydro-gravitational-dynamics theory HGD, gravitational structure formation begins 30,000 years after the turbulent big bang by fragmentation into super-cluster-voids and super-clusters. Proto-galaxies in linear and spiral clusters are the smallest fragments to emerge from the plasma epoch at decoupling at $10^{13}$ s with a turbulent morphology determined by the plasma turbulence and the Nomura scale $10^{20}$ m, which is determined by gravity, the fossilized density and rate-of-strain of the $10^{12}$ s time of first structure and the large photon viscosity of the plasma. After decoupling, the gas proto-galaxies fragment into $10^{36}$ kg proto-globular-star-cluster PGC clumps of earth-mass $10^{25}$ kg hot gas clouds that eventually freeze to form primordial-fog-particle PFP dark matter planets. This is the galaxy dark matter BDM. The non-baryonic-dark-matter NBDM is so weakly collisional that it fragments only after decoupling to form galaxy cluster halos. It does not guide galaxy formation according to the cold-dark-matter hierarchical clustering CDMHC theory, which is fluid mechanically untenable and obsolete. NBDM has 97% of the mass of the universe and serves to bind together rotating clusters of galaxies by gravitational forces. The rotation of galaxies reflects density gradients of big bang turbulent mixing and the sonic expansion of proto-super-cluster-voids. The spin axis appears for low wavenumber spherical harmonic components of CMB temperature anomalies, the Milky Way and galaxies of the local group, and extends to $4.5 \times 10^{25}$ m (1.5 Gpc) in quasar polarization vectors, supporting a big bang turbulence origin. Gas proto-galaxies stick together by frictional processes of the frozen gas planets, just as PGCs have been meta-stable for the 13.7 Gyr age of the universe. Evidence of PGC-friction is inferred from the local group Hubble diagram and from redshift anomalies of Hickson compact galaxy groups such as the Stephan Quintet compared to Sloan Digital Sky Survey SDSS galaxy observations.

## INTRODUCTION

The standard model of cosmology is in the process of rapid decomposition as a relentless flood of new data confronts old ideas [1,2]. New space telescopes cover an ever-widening band of frequencies. Ground based telescopes are linked and controlled by ever more powerful computers that track events as they happen and freely distribute the information nearly real time on the internet. This standard ("concordance") model of cosmology is cold-dark-matter hierarchical-clustering CDMHC based on an acoustic length scale proposed in 1902 by Jeans [1] as the single (but fluid mechanically untenable) criterion for gravitational structure formation (see Table 1). In 1902 only the Milky Way nebula of stars was recognized as a galaxy. As pointed out by Hoyle, Burbidge and Narlikar 2000 [3] spiral nebula (galaxy) Messier 51 had been detected in 1855 by Lord Rosse, and it was then only speculated that such objects were Milk-Way-like galaxies, a view strongly dismissed by Agnes Clerke in her 1905 well-known popular book *The System of Stars* based on her perception of the results and conclusions of professional astronomers of the day:

> *The question whether nebulae are external galaxies hardly any longer needs discussion. It has been answered by the progress of research. No competent thinker, with the whole of the available evidence before him, can now, it is safe to say, maintain any single nebula to be a star system of co-ordinate rank with the Milky Way. A practical certainty has been attained that the entire contents, stellar and nebula, of the sphere belong to one mighty aggregation, and stand in ordered mutual relations within the limits of one all embracing scheme [4].*

As Hoyle et al. note in their preface, pressures of big science funding have badly corrupted the peer review system of astrophysics and cosmology. Papers that deviate from the standard model in any way are likely to be dismissed out of hand by referees and scientific editors that depend on big science funds for survival. It was pointed out in 1996 [10] that the Jeans 1902 fluid mechanical analysis of standard cosmology is fatally flawed and obsolete from the neglect of various basic principles of fluid mechanics termed hydro-gravitational-dynamics HGD [7-23], but publication of this information has so far not been allowed in any "reputable" astrophysical journal. Contrary to the linear instability criterion of Jeans 1902, gravitational instability is highly non-linear, fluid mechanically limited and absolute [13]. It is easy to show that viscous and turbulent forces are critically important and that diffusivity of the nearly collision-less non-baryonic-dark-matter NBDM prevents Jeans condensation and hierarchical clustering of CDM halos during the plasma epoch before decoupling. Artificial "Plummer forces" etc. introduced to fit data from observations by numerical simulations (see Table 1 and Dehnen 2001) compensate for the physical impossibility of CDM halo formation and clustering [5,6]. The "Plummer force length scales" required to permit numerical simulations to match super-void observations [31] match the Nomura scale $L_N = 10^{20}$ m, reflecting proto-globular-star-cluster PGC friction from planets-in-clumps as the dominant form of galaxy dark matter, as inferred from quasar microlensing observations by Schild in 1996 [24].

Proto-galaxy-clusters form at the last stage of the plasma epoch guided by weak turbulence along vortex lines produced by expanding proto-supercluster-voids encountering fossil density gradients of big bang turbulence [16,17] producing baroclinic torques and turbulence on the expanding void boundaries. In this paper we focus on the evolution of proto-galaxy-clusters during the present gas epoch. Examples of gas proto-galaxy-clusters are shown in Figure 1, from the Hubble Space Telescope Advanced Camera for Surveys HST/ACS observing the Tadpole galaxy merger UGC 10214. We see that these dimmest objects (magnitude 24-28) with $z > 0.5$ clearly reflect their formation along turbulent vortex lines of the plasma epoch, and clearly reflect the gentle nature of the early universe as these gas proto-galaxy-clusters expand ballistically and with the expansion of the universe against frictional forces of the baryonic dark matter. The mechanism of momentum transfer between gas proto-galaxies is much better described by the 1889 meteoroid collision mechanism of G. H. Darwin [2] than by the 1902 collisionless-gas mechanism of J. H. Jeans [1].

From HGD, the chains of star clumps shown in Fig. 1 have been incorrectly identified as "chain galaxies" rather than chains of proto-galaxies since their discovery in the Hawaiian Deep Field by Cowie et al. 1995 at magnitude 25-26. The rows of clumps are not edge-on spirals and the tadpoles are not end-on chain galaxies as suggested by Elmegreen et al. 2004 [39]; they are remnants of gravitationally produced plasma fragments along turbulent vortex lines of the primordial plasma produced at the Nomura scale of plasma proto-galaxies. In the following we show end-on proto-galaxy-clusters can best be explained as the "fingers of God" structures (Fig. 8 top) observed in the Sloan Digital Sky Survey II and by the Hickson 1993 Compact Group HCG class of galaxy clusters [29] exemplified by Stephan's Quintet [22], Figure 2 (Fig. 8 bottom). A complex system of star wakes, globular star cluster wakes and dust trails leaves little room to doubt that the dark matter ha-los of the SQ galaxy system are dominated by PGC clumps of planets from which the stars and GCs form on agitation, and that the SQ galaxies have frictionally resisted Hubble flow separation starting from a primordial plasma chain-proto-galaxy-cluster.

The linear gas-proto-galaxy clusters of Fig. 1 show that the universe soon after decoupling must have been quite gentle for them to survive. This contradicts CDM models of galaxy formation where the first galaxies are CDM haloes that have grown by hierarchical clustering to about $10^{36}$ kg (a globular cluster mass) and collected a super-star amount of gas in CDM-halo gravitational potential wells, about $10^{32}$ kg (100 solar mass). As soon as the gas cools sufficiently so that its Jeans scale permits condensation it does so to produce one super-star and one extremely bright supernova.

The combined effect of these (fictional) mini-galaxy supernovae of Population III stars is so powerful that the entire universe of gas is re-ionized according to CDMHC. The problem with this scenario is that it never happened. It rules out the formation of old globular star clusters that require very gentle gas motions. The extreme brightness of the first light is not observed [19]. Re-ionization is not necessary to explain why neutral gas is not observed in quasar UV-spectra once it is understood that the dark matter of galaxies is frozen PFP primordial planets in PGC clumps.

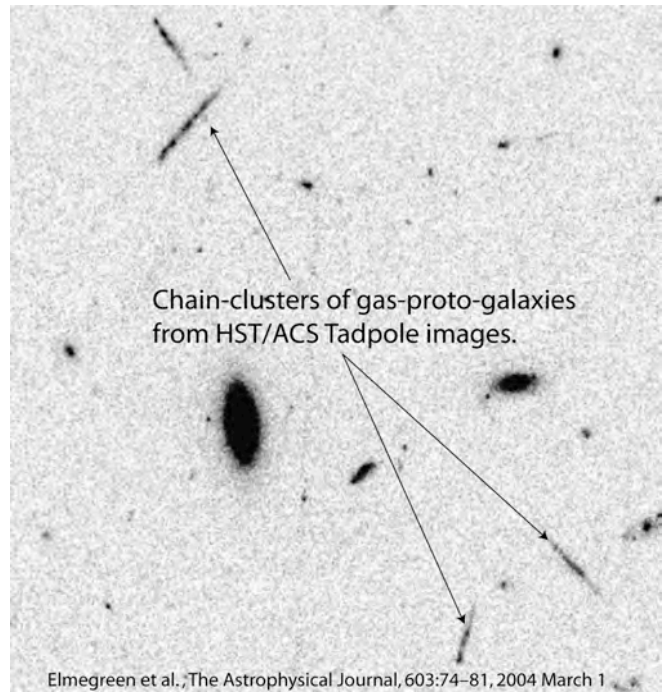

Figure 1. Chain-clusters of gas-proto-galaxies GPGs reflect their origin by gravitational fragmentation along turbulent vortex lines of the plasma epoch at decoupling time t ~ $10^{13}$ s (300,000 years) after the big bang. Only ~0.1% of the baryonic dark matter BDM (planets in clumps) has formed old-globular-star-cluster OGC small stars in these proto-galaxies. More than 80% of the dimmest proto-galaxies (magnitudes 24-28) are in linear proto-galaxy-clusters termed chains, doubles and tadpoles. The tadpole tails and the luminosity between all GPGs are stars formed from frictional BDM planets that form gas when agitated and accrete to form stars.

The chain-clusters of GPGs in Fig. 1 confirm the prediction of HGD that the early universe was quite gentle compared to that required by CDM and that the proto-galaxies start their evolution in the gas epoch at very small scales compared to galaxy sizes observed today. From HGD, each GPG in the linear clusters has about $10^{42}$ kg of BDM dark matter PGCs. The foreground elliptical and spiral galaxies shown in Fig. 1 have not acquired mass by merging according to CDMHC. The BDM gradually diffuses out of the $10^{20}$ m proto-galaxy core to form the $10^{22}$ m BDM halo against PGC frictional forces. The PGC frictional forces trigger the formation of BDM halo star trails and inhibit the ballistic and Hubble flow growth, as seen in end-on images of linear gas-galaxy-clusters.

From HGD, gas-proto-galaxies soon after formation are quite sticky from PGC friction. Even though plasma-proto-galaxies are stretching apart along turbulent vortex lines with the maximum rate of strain of the turbulence the central GPGs formed may not be able to separate. An example is shown in Figure 2, the spectacular Stephan's Quintet SQ. Three galaxies in a narrow range of angles have precisely the same redshift $z = 0.022$, one has $z = 0.019$ but one has a highly anomalous $z = 0.0027$. SQ is one of many highly compact galaxy clusters termed Hickson Compact Groups HCG. Nearly half of the HCG galaxy clusters have at least one highly anomalous redshift member.

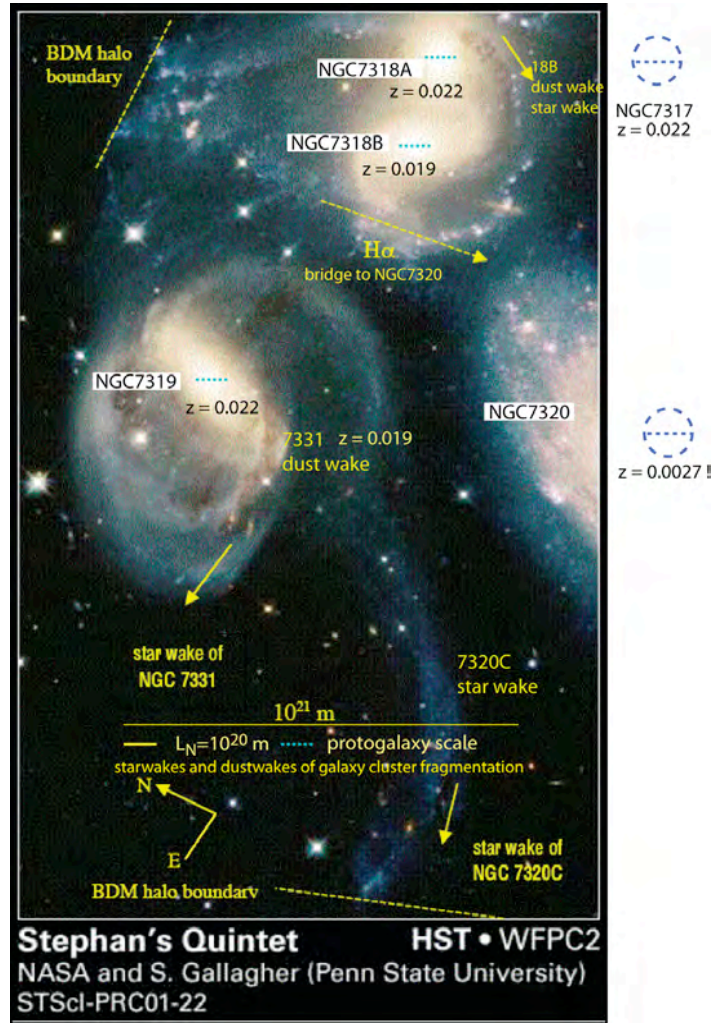

Figure 2. Stephan's Quintet (SQ, HCG 92, Arp 319, VV 288) very compact and mysterious galaxy cluster (Stephan 1877). The Trio NGC7319, NGC7318A, NGC7317 have redshift 0.022, NGC 7318B with redshift 0.019 that matches that of NGC7331, but nearby NGC7320 at anomalous redshift 0.0027. All are connected by star-trails, star-cluster-trails and dust-trails, suggesting an end-on chain-cluster of proto-galaxies formed along a turbulent vortex line [22]. Note the sharp BDM halo boundaries (dashed lines) where star formation ceases as the galaxies separate.

Burbidge and Burbidge 1961 measure and report the anomalous nature of the SQ galaxy redshifts. For virial equilibrium the closely aligned NGC 7313AB galaxies require an M/L ratio of 300 from dynamical models, but NGC 7320 requires M/L of 10,000, much too large to be credible. A connecting gas bridge to NGC 7320 (Fig. 2) proves it is not a chance intruder but a separated companion [22]. One suggested possibility is that galaxies are ejected by AGN parent galaxies with intrinsic redshifts [3,33,34,35], which accounts for the Burbidge 2003 observational fact [33] that AGN galaxies (eg. NGC 6212) are observed with many more nearby QSOs than chance allows. An alternative interpretation from HGD is that Hickson Compact Groups and SQ are simply end-on views of chain-galaxy-clusters (Fig. 1), and distant quasar (QSO) galaxies are often included in end-on views of linear clusters (Fig. 8) because they are bright. The Trio is still stuck together by PGC friction and 7313B and 7320 have separated ballistically and from the expansion of the universe. Their close angular proximity is an optical illusion due to their nearness to earth and perspective.

From known properties of the hot big bang universe the Schwarz viscous and turbulent scales of Table 1 show fragmentation will occur early at massive proto-super-cluster scales by formation of expanding super-cluster-voids independent of the NBDM.

Table 1. Length scales of gravitational instability

| Length Scale Name | Definition | Physical Significance |
| --- | --- | --- |
| Jeans Acoustic | $L_J = V_S/(\rho G)^{1/2}$ | Acoustic time matches free fall time |
| Schwarz Viscous | $L_{SV} = (\gamma \nu/\rho G)^{1/2}$ | Viscous forces match gravitational forces |
| Schwarz Turbulent | $L_{ST} = (\varepsilon/[\rho G]^{3/2})^{1/2}$ | Turbulent forces match gravitational forces |
| Schwarz Diffusive | $L_{SD} = (D^2/\rho G)^{1/4}$ | Diffusive speed matches free fall speed |
| Horizon, causal connection | $L_H = ct$ | Range of possible gravitational interaction |
| Plummer force scale | $L_{CDM}$ | Artificial numerical CDM halo sticking length |

$V_S$ is sound speed, $\rho$ is density, G is Newton's constant, $\gamma$ is the rate of strain, $\nu$ is the kinematic viscosity, $\varepsilon$ is the viscous dissipation rate, D is the diffusivity, $c$ is light speed, $t$ is time.

The plan of the present paper is to first review the very different predictions of HGD and CDMHC theories with respect to galaxy formation and evolution. Then Observations are discussed, followed by a Conclusion.

**THEORY**

Figure 3 shows the sequence of gravitational structure formation events according to hydro-gravitational-dynamics HGD cosmology leading to primordial gas-proto-galaxies and evolved galaxies of the present time. A hot big bang is assumed at 13.7 Gyr before the present, followed by an inflation event where big bang turbulent temperature microstructure is fossilized by stretching of space beyond the scale of causal connection $L_H = ct$, where $c$ is the speed of light and $t$ is the time. Gravitational instability produces the first structure in the plasma epoch by fragmentation, where proto-supercluster-voids begin to grow at $10^{12}$ seconds (30,000 years) leaving proto-superclusters in between. The proto-super-clusters do not collapse by gravity but expand with the expansion of space working against plasma photon viscosity. Viscous dissipation rates estimated from $\varepsilon \sim \nu\gamma^2$ give $\varepsilon \sim 400$ m$^2$ s$^{-1}$. Photon-electron collision lengths were $\sim 10^{18}$ m, less than the horizon scale $L_H = 3 \times 10^{20}$ m as required by continuum mechanics. Viscous dissipation rates in the gas epoch after decoupling decreased with $\nu$ and $\gamma$ to $\varepsilon \sim 10^{-13}$ m$^2$ s$^{-1}$, small enough (Table 1) to permit formation of dark matter planets in clumps and the small stars of old globular star clusters.

The HGD time for the first plasma gravitational fragmentation is when the Schwarz viscous scale $L_{SV}$ is less than $L_H$ (see Table 1). Proto-super-voids grow as rarefaction waves at sound speed $c/3^{1/2}$. The NBDM decouples from the plasma because it is weakly collisional. Turbulence is produced at expanding void boundaries by baroclinic torques. Observations of Sreenivasan and Bershadskii [40,41,42] confirm that the Reynolds number of the turbulence is rather weak. Fragmentations and void formations occur at smaller scales until the plasma to gas transition (decoupling) at $10^{13}$ seconds (300,000 years). The weak turbulence produces plasma-proto-galaxies by fragmentation, with NBDM filled voids formed along stretching and spinning turbulent vortex lines.

In Fig. 3, a. Cosmic Microwave Background temperature anisotropies reflect structures formed in the plasma epoch. b. From HGD the photon viscosity of the plasma epoch prevents turbulence until the viscous Schwarz scale $L_{SV}$ becomes less than the Hubble scale (horizon scale, scale of causal connection) $L_H = ct$, where $c$ is the speed of light and $t$ is the time. The first plasma structures were proto-super-cluster voids and proto-super-clusters at $10^{12}$ seconds (30,000 years). c. Looking back in space is looking back in time. Proto-galaxies were the last fragmentations of the plasma (orange circles with green halos) at $10^{13}$ seconds. d. The scale of the gravitational structure epoch is only $3 \times 10^{21}$ m compared to present supercluster sizes of $10^{24}$ m and the largest observed super-void scales of $10^{25}$ m (Fig. 5). e. Turbulence in the plasma epoch is generated by baroclinic torques on the boundaries of the expanding super-voids.

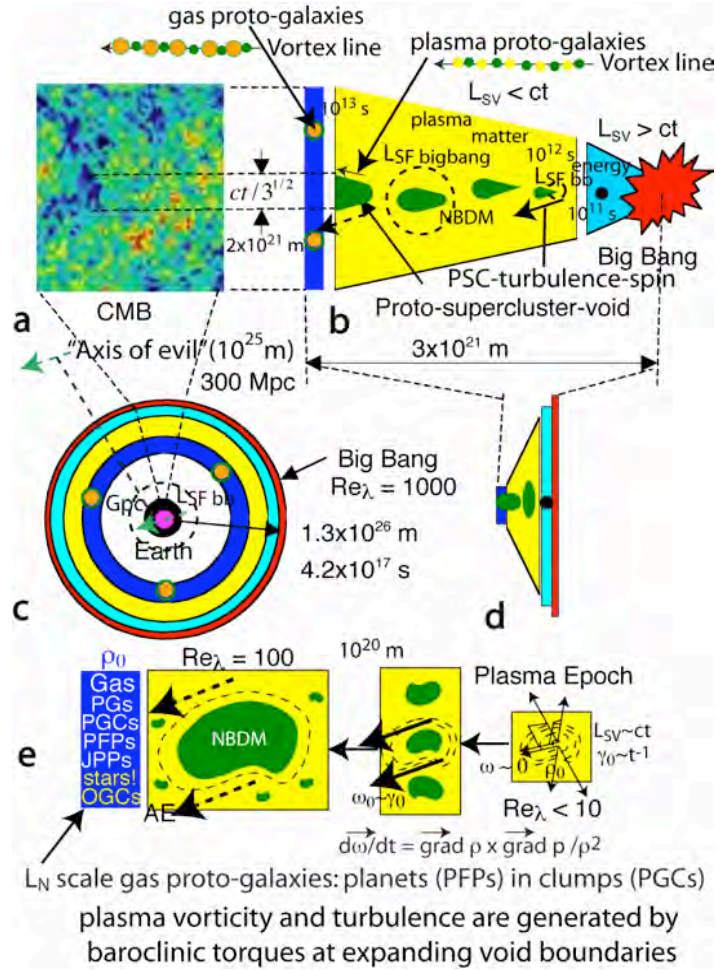

Figure 3. Protogalaxy formation at the end of the plasma epoch by hydro-gravitational-dynamics HGD theory. Evidence of spin to more than Gpc scales has been detected [43].

Gas chains of proto-galaxies are formed at decoupling, as shown by the two schematic diagrams at the top of Fig. 3a. Because photons suddenly decouple from electrons the viscosity of the fluid suddenly decreases by a factor ~ $10^{13}$, greatly decreasing the viscous Schwarz scale $L_{SV}$ and the Jeans scale $L_J$ fragmentation masses. The two fragmentation scales work simultaneously with the same gravitational free fall time to produce Jeans-mass clumps PGCs of earth-mass gas planets PFPs, which today constitute the baryonic-dark-matter BDM halos of galaxies.

Figure 4 illustrates super-void and galaxy formation following the standard cosmological model (ΛCDMHC) advocated in the Peebles 1993 book *Principles of Physical Cosmology*. According to the Peebles 1993 timetable (Table 25.1, p 611) [25] super-clusters, walls and voids form at redshift $z \sim 1$; that is, at least 5 Gyr after the big bang versus 0.0003 Gyr from Gibson 1996 [14] and HGD. But completely empty super-voids (see Fig. 5) have been detected by radio telescopes with sizes at least 300 Mpc or $10^{25}$ m, 10% of the horizon scale $L_H = 10^{26}$ m [30]. Peebles 2007 [26] recognizes that observations of empty voids even on smaller locally observed scales of $10^{24}$ m is a possibly insuperable problem for CDMHC models.

Fig. 4 contrasts the predictions of ΛCDMHC theory of galaxy and void formation with observations and the predictions of HGD theory. Failures are indicated by red Xs. HGD cosmology is driven by turbulent combustion at Planck scales from Planck-Kerr instability, with Taylor microscale Reynolds number $Re_\lambda \sim 1000$. Gluon viscosity terminates the event after cooling from $10^{32}$ K Planck temperatures to $10^{28}$ K strong force freeze-out temperatures where quarks and gluons can appear. Turbulent temperature patterns are frozen as turbulence fossils by exponential inflation of space driven by negative stresses of both turbulence and gluon viscosity pulling $10^{90}$ kg of mass-energy out of the vacuum against the Planck tension $c^4/G$. The mass-energy of our present horizon is only $\sim 10^{53}$ kg, so we can see only $\sim 10^{-40}$ fraction of the universe produced by the big bang. A black dot in the blue inflation triangle of Fig. 3 symbolizes the $\sim 20\%$ temperature fluctuations expected from big bang turbulence, contrasting with tiny quantum-mechanical fluctuations expected in the standard model in Fig. 4.

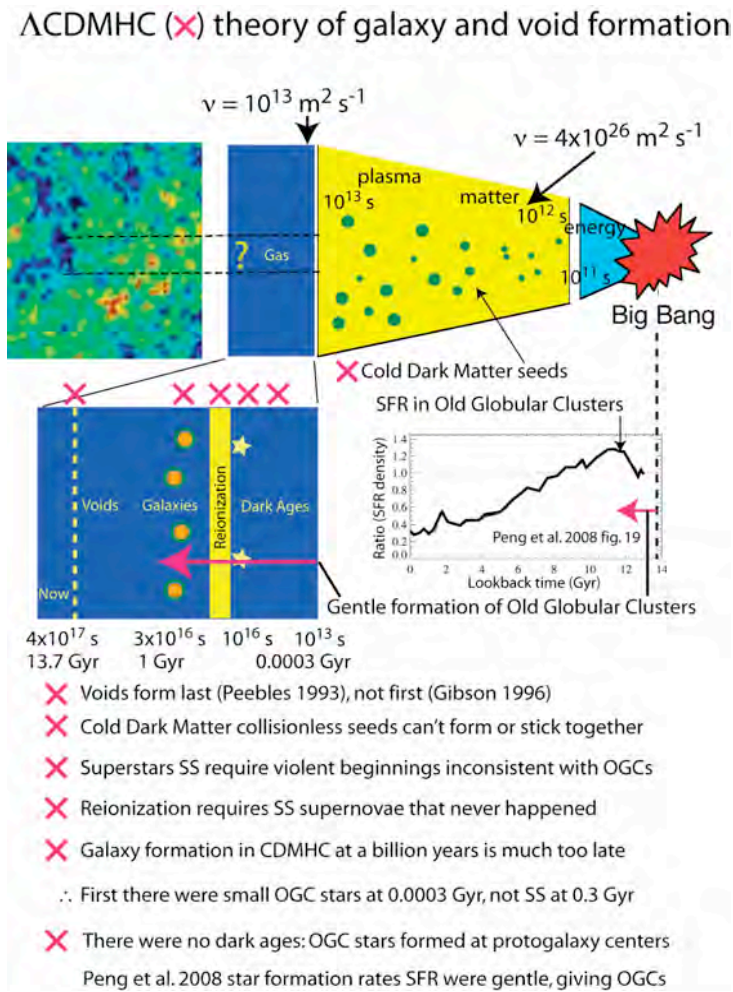

Figure 4. Void and galaxy formation by the standard ΛCDM theory are impossible to reconcile with observations and fluid mechanical HGD theory (see text). Several failed aspects of ΛCDMHC theory are indicated by red Xs. It is fatally flawed and must be abandoned.

In CDMHC models voids form last rather than first, so this difference is most easily tested by observations. The time of first void formation from HGD is $10^{12}$ s, compared to ~ $10^{17}$ s for CDMHC. Star formation rates of Peng et al. 2008 [27] favor small old globular-cluster stars. These could not possibly be formed under the violent conditions of galaxy formation and mergers intrinsic to CDMHC.

We now examine available observations for comparison with theories of galaxy formation and evolution.

**OBSERVATIONS**

Evidence of the large primordial super-voids of HGD theory (Fig. 1) is shown in Figure 5, from Rudick et al. 2008. Focusing on the direction of the anomalous "cold spot" of the CMB it was found that a $10^{25}$ m (300 Mpc) completely empty region could explain the ~ $7 \times 10^{-5}$ °K CMB cold spot by the integrated Sachs-Wolfe method. The empty region is estimated to be at redshifts $z \sim 1$, and is therefore completely impossible to explain by CDMHC models where super-voids are formed last rather than first (Fig. 4). The probability of such a void forming from concordance CDM models is estimated [30] to be $< 10^{-10}$.

Tinker and Conroy 2008 [31] explain the void phenomenon by numerical simulations and produce a relatively empty super-void of scale $10^{24}$ m using numerical simulations and numerically convenient but entirely imaginary "Plummer forces" [6]. As mentioned previously, these Plummer etc. forces are physically untenable for weakly collisional NBDM materials such as CDM and cannot possibly explain the $10^{25}$ m void of Fig. 5.

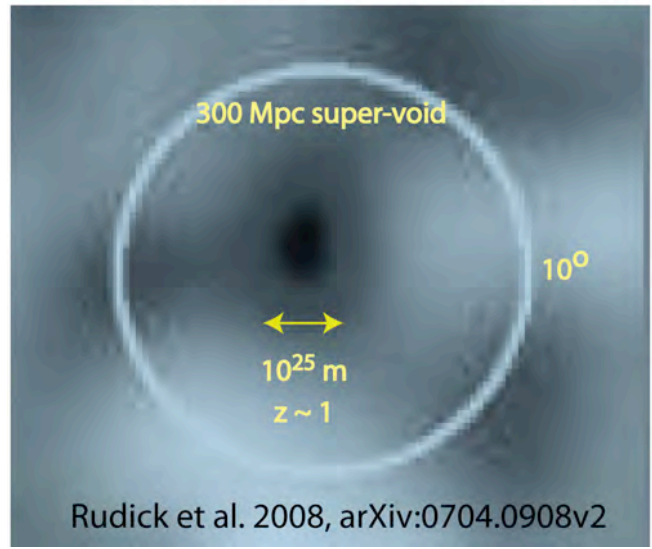

Figure 5. Super-void detection by the radio telescope very large aperture (NVSS) survey in the direction of the anomalous cold spot of the CMB. Such large voids are expected from HGD but are quite impossible using CDM models [30].

Figure 6 summarizes observational evidence that the dimming of supernova Ia events is not evidence for dark energy and a cosmological constant $\Lambda$, but is merely a systematic error due to the presence in the near vicinity of shrinking white dwarf stars approaching the Chandrasekhar instability limit (1.44 $M_{Sun}$) of BDM frozen planets partially evaporated by strong spin-powered plasma jets. Dimness of the SNe Ia events increases with their magnitudes, and cannot be explained by uniform grey dust as shown by the top curve using estimates of Reiss et al. 2004 [36].

In Fig. 6, open circles emphasize SNe Ia events un-obscured by evaporated BDM planet atmospheres (no dark energy) for the Reiss et al. dimness models. Solid red ovals emphasize events partially obscured by planet atmospheres (non-linear grey dust). Thousands of BDM planets (right insert) in the Helix planetary nebula are evaporated by spin-powered plasma jets and radiation from the central white dwarf. From HGD the Helix PNe is not ejected from a massive precursor; instead, the BDM planets are evaporated in place. The observed dimness requires post-turbulent electron-density fluctuations in gas with density $\rho \sim 10^{-12}$ kg m$^{-3}$ sufficient to permit turbulence. The large 20-30% dimness at $z \sim 0.5$ [36] cannot be explained by reasonable quantities of dust or gases alone in the observed $10^{13}$ m planet atmospheres shown in the Helix PNe BDM planets insert: it requires fossil electron-density turbulence (post-turbulent-microstructure) forward scattering [44].

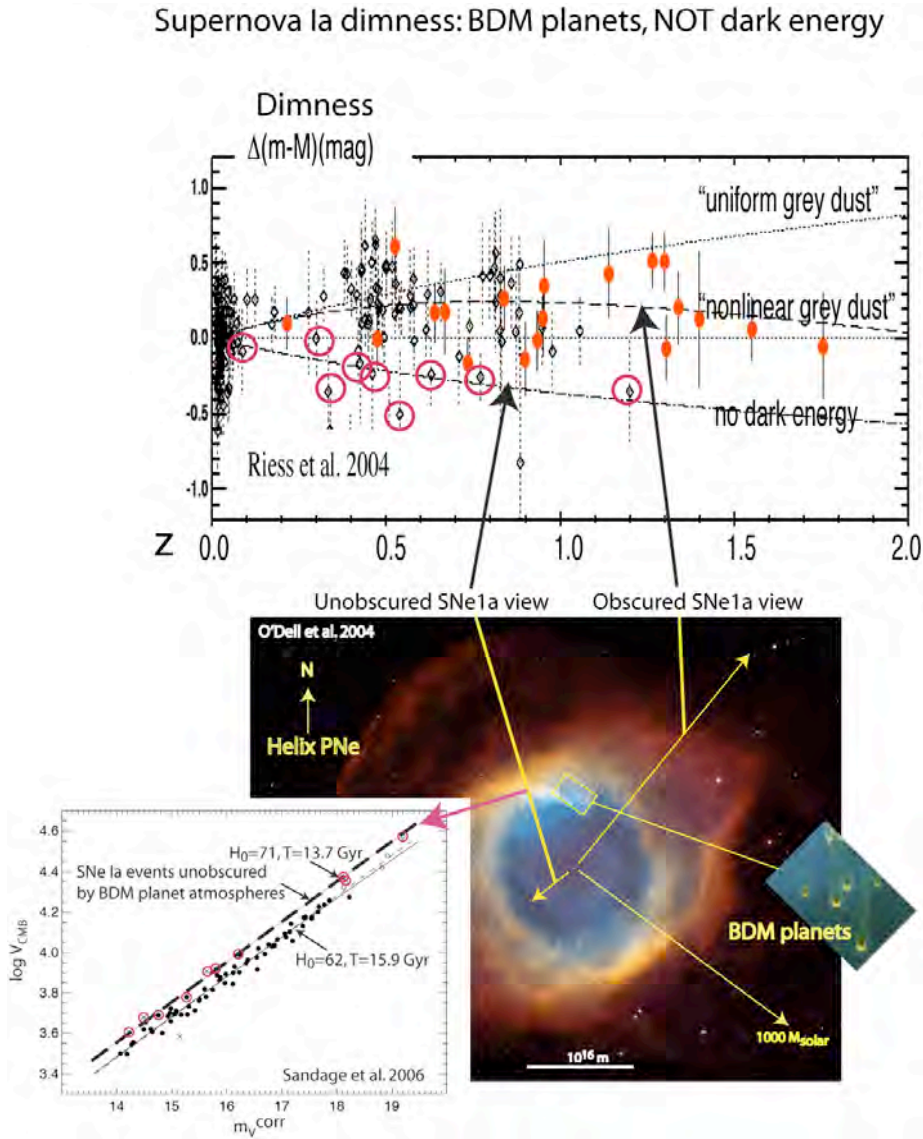

Figure 6. Observations that show the anomalous dimness of supernova Ia events of Reiss et al. 2004 and the anomalously low Hubble constants of Sandage et al. 2006 [37] can be attributed to BDM planet atmospheres, not dark energy [20]. The nearby Helix planetary nebula PNe at $6\times10^{18}$ m has a central white dwarf with polar jet that evaporates ambient BDM planets of its PGC. A close-up view of the evaporated planets is shown in the insert on the right.

From HGD, all stars are formed from BDM planets in PGCs. All PGCs have the primordial density $\rho_0 \sim 4 \times 10^{-17}$ kg m$^{-3}$, which matches the density of globular star clusters. The $10^{13}$ m size of the planet atmospheres with $10^{14}$ m separation observed are consistent with this primordial density [20].

The Fig. 6 insert at lower left shows the Sandage et al. 2006 SNe Ia study of the Hubble Constant $H_0$, carefully corrected for Cepheid variable distances and locations. The age of the universe is 15.9 Gyr from this study, which is unacceptably large. However, open red circles show SNe Ia event lines of sight un-obscured by BDM planet atmospheres from HGD. These agree with the CMB age of the universe of 13.7 Gyr. Gamma ray burst dimnesses support this interpretation [38].

Figure 7 shows velocities $V_{LG}$ in km s-1 of the local group of galaxies as a function of their distances in Mpc so the slope of a line from the origin is a measure of the Hubble Constant.

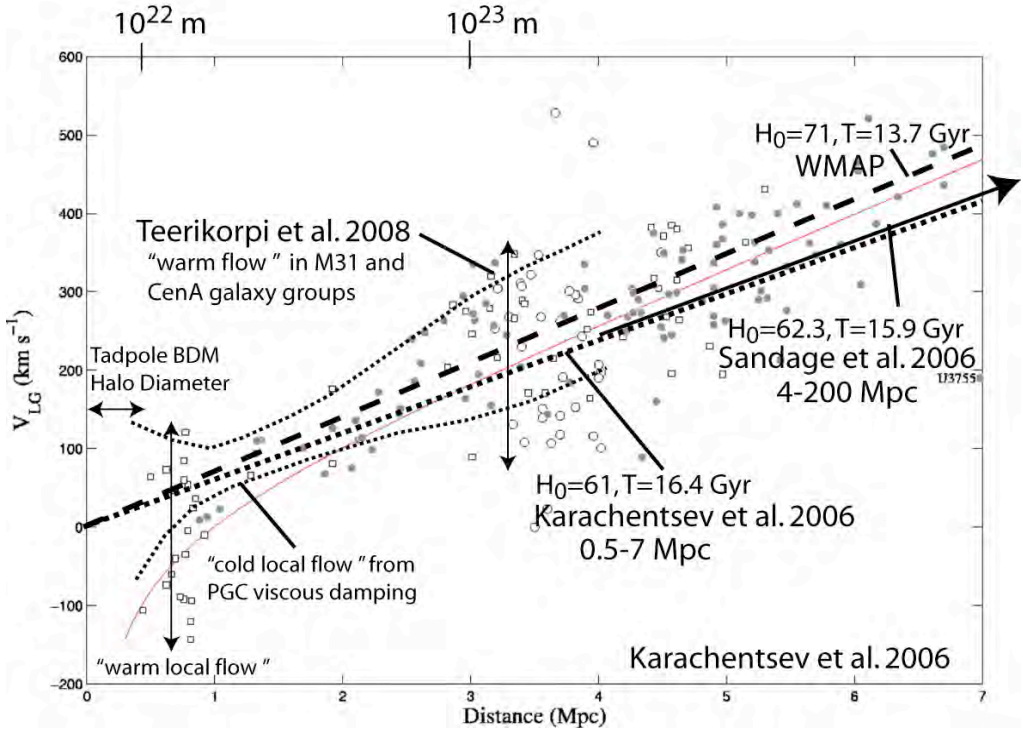

Figure 7.  Estimates of the Hubble Constant for galaxies in the local group show wide scatter out to distances of a Mpc due to frictional interactions of BDM halos. The Tadpole BDM halo size is shown by the horizontal double arrow [23]. Beyond this distance the galaxies begin to separate due to Hubble flow. Warm flows in M31 and CenA galaxy groups have $H_0$ dispersions similar to that of galaxies near the Milky Way, as shown by vertical double arrows. The dotted line is an extrapolation to the Sandage et al. 2006 Hubble constants shown in Fig. 6, at 4 to 200 Mpc.

Figure 8 shows a Sloan Digital Sky Survey map of local galaxies compared to the HGD interpretation of Stephan's Quintet as an end-on chain of gas proto-galaxies. In the top panel, note that galaxies with gently formed old stars from HGD, indicated by red dots, are often aligned in thin pencils termed "fingers of god". Blue dots denote galaxies with more blue stars, indicating stronger tidal agitation and large $L_{ST}$ values from HGD. The HGD interpretation is that the SDSS II galaxies are relatively near to earth, with redshift $z \sim 0.1$ or less, so perspective causes a decrease in angular separation for distant galaxies that are already nearly aligned. An arrow shows $10^{25}$ m, about 10% of the present horizon $L_H$. The red pencil-like features are interpreted from HGD as chain clusters of old galaxies aligned by vortex lines in the plasma epoch that have continued moving along these directions ballistically and from the Hubble flow homogeneous straining of the universe during the gas epoch. Dashed circles indicate the local $10^{24}$ m (30 Mpc) super-void and super-cluster scales.

PGC frictional stickiness has inhibited separation of the galaxies along their axes and in transverse directions, as shown by the dust, GC, and star trails of Stephan's Quintet (Fig. 2 and Fig. 8 bottom).

Fig. 8a summarizes the history of SQ formation starting from the time of first fragmentation to plasma to gas transition.  Fig. 8b shows SQ at present, with the Trio about $4 \times 10^{24}$ m distant and NGC 7320 with redshift 0.0027 at about $10^{23}$ m.  A cartoon of the SQ galaxies is shown near the origin of the SDSS II Galaxy Map.

It seems clear from Fig. 8 that the Trio of SQ galaxies are not clustered by chance or by merging but were formed simultaneously in a linear cluster of proto-galaxies along a turbulent vortex line of the plasma epoch (Fig. 1a).  For 13.7 Gyr they have resisted separation by the expansion of the universe due to PGC friction from frictional interactions of galaxy BDM dark matter planets in galaxy dark matter halos [22].

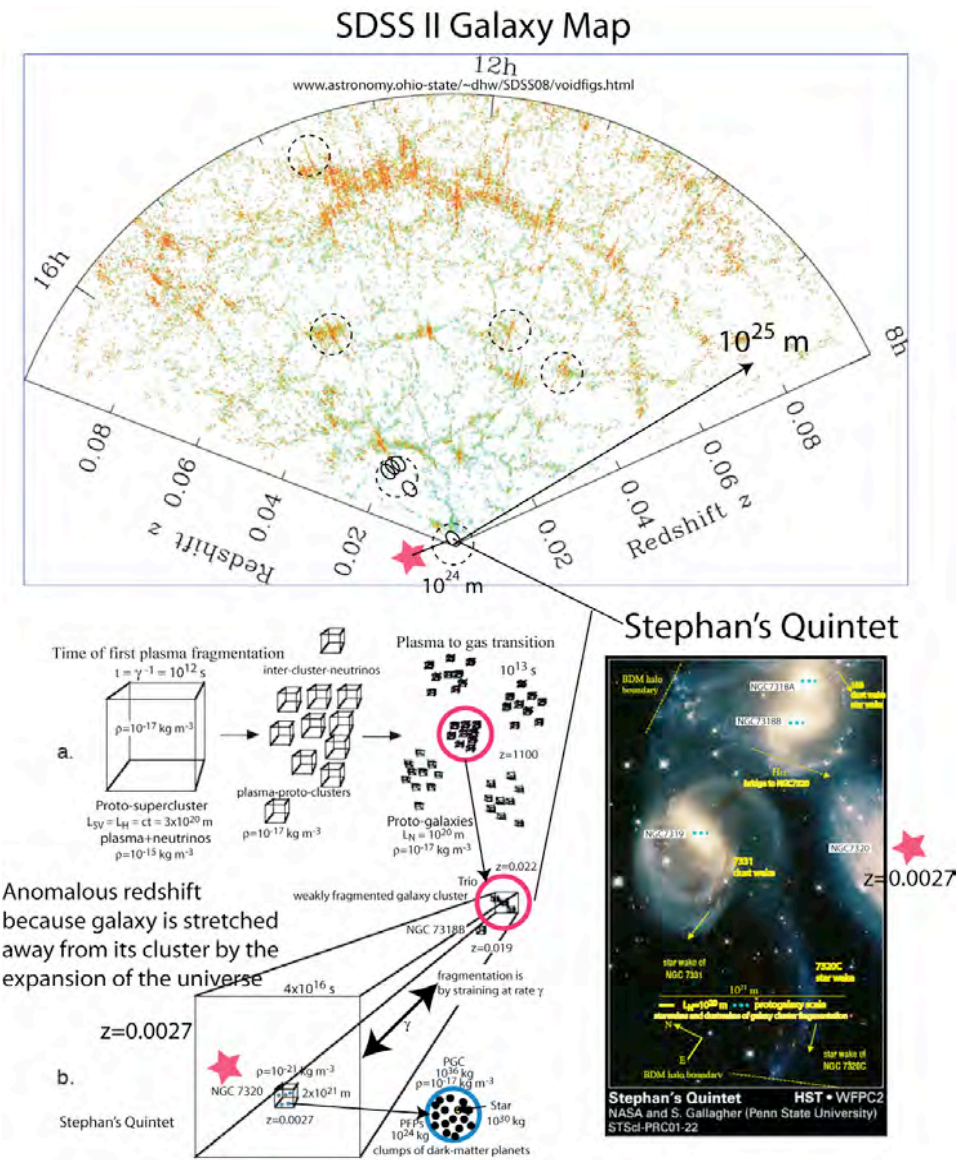

Figure 8.  Stephan's Quintet provides evidence that linear galaxy clusters were formed by fragmentation in the plasma epoch along turbulent vortex lines [22].  Red stars indicate the most anomalous galaxy NGC 7320 with redshift $z = 0.0027$ in three views.

The HGD interpretation of Fig. 8 is that PGC friction inhibits the separation of galaxies in the gas epoch by collisional and tidal interactions of BDM planet halos.  The Arp 1973 suggestion that NGC 7331 has ejected the other galaxies with intrinsic redshifts is unnecessary, and would require introduction of an unknown class of new physical laws.

## CONCLUSION

Observations exclude the ΛCDMHC standard model for galaxy and void formation as the last steps of gravitational structure formation rather than the first from HGD where the Jeans 1902 theory that provides the basis of CDM is obsolete and fluid mechanically untenable. HGD explains the formation of galaxies as the last stage of gravitational fragmentation starting early in the plasma epoch with proto-super-clusters and proto-super-voids, and finishing with plasma-proto-galaxy morphology determined by weak turbulence from gravitational void expansions and fossil turbulence density gradients from the epoch of strong big bang turbulence. The evolution of gas proto-galaxies from HGD is extremely gentle compared to an unnecessarily violent epoch of super-star formation, supernovae and re-ionization required by ΛCDMHC. Population III stars and these associated CDM events never happened.

Dimming of SNe Ia events by evaporated BDM planet atmospheres provides an HGD alternative to the new physical laws required by the dark energy hypothesis, Λ, and the Sandage et al. 2006 evidence that the universe age is 15.9 Gyr. There is no dark energy, there is no Λ, and corrections for dimming give a universe age of 13.7 Gyr.

Stephan's Quintet confirms predictions of HGD about the evolution of chain gas-proto-galaxy clusters and the importance of PGC friction to stick proto-galaxies together and resist ballistic forces and universe-space-expansion that try to move them apart. The interpretation of SQ and chain-galaxy-clusters by HGD theory provides an alternative to suggestions [3,33,34,35] that central galaxies in chain clusters can emit galaxies and quasars with intrinsic redshifts. Globular cluster wakes, star wakes and dust wakes clearly show the galaxies of SQ were formed in a linear chain and have separated, never merged, as the galaxy cluster has evolved against the PGC friction of the galaxy-dark-matter-halos consisting of frozen planets in GC-mass clumps.

Further evidence of proto-globular-star-cluster friction from dark-matter-planet interactions is provided by the Hubble diagrams of Fig. 7 for the local group and Fig. 6 from Sandage et al. 2006 at larger distances up to 200 Mpc from SN Ia. PGC-friction explains the random scatter of galaxy velocities in clusters for ~ Mpc lengths scales small enough for BDM halos to interact. At larger scales the expansion of the universe becomes the dominant mechanism to separate galaxies.